\documentclass[twocolumn]{aastex631}

\usepackage{comment}

\newcommand{\lya}{Ly$\alpha$}
\newcommand{\fcgs}{erg~s$^{-1}$~cm$^{-2}$~\AA$^{-1}$}

\newcommand{\flcgs}{erg~s$^{-1}$~cm$^{-2}$}
\newcommand{\sblcgs}{erg~s$^{-1}$~cm$^{-2}$~arcsec$^{-2}$}

\newcommand{\nlae}{19}
\newcommand{\overdens}{25}

\shorttitle{A $z\approx 4$ filament in the MUDF}
\shortauthors{Tornotti et al.}

\begin{document}

\title{The MUSE Ultra Deep Field: A 5~Mpc stretch of the $z\approx 4$ cosmic web revealed in emission}


\shorttitle{A 5-Mpc-stretch of the $z\approx 4$ cosmic web.}

\author[0009-0001-3388-8742]{Davide Tornotti}
\affiliation{
Dipartimento di Fisica ``G. Occhialini'', 
Universit\`a degli Studi di Milano-Bicocca,
Piazza della Scienza 3, I-20126 Milano, Italy}

\author[0000-0001-6676-3842]{Michele Fumagalli}
\affiliation{
Dipartimento di Fisica ``G. Occhialini'', 
Universit\`a degli Studi di Milano-Bicocca,
Piazza della Scienza 3, I-20126 Milano, Italy}
\affiliation{INAF – Osservatorio Astronomico di Trieste,
Via G. B. Tiepolo 11,
I-34143 Trieste, Italy}

\author[0000-0002-9043-8764]{Matteo Fossati}
\affiliation{
Dipartimento di Fisica ``G. Occhialini'', 
Universit\`a degli Studi di Milano-Bicocca, 
Piazza della Scienza 3, I-20126 Milano, Italy}
\affiliation{INAF – Osservatorio Astronomico di Brera, 
Via Brera 28,
I-21021 Milano, Italy}

\author[0000-0002-4770-6137]{Fabrizio Arrigoni Battaia}
\affiliation{Max-Planck-Institut f\"ur Astrophysik, 
Karl-Schwarzschild-Str. 1, D-85748 Garching bei M\"unchen, Germany}

\author[0000-0001-8261-2796]{Alejandro Benitez-Llambay}
\affiliation{
Dipartimento di Fisica ``G. Occhialini'', 
Universit\`a degli Studi di Milano-Bicocca, 
Piazza della Scienza 3, I-20126 Milano, Italy}

\author[0000-0001-8460-1564]{Pratika Dayal}
\affiliation{Kapteyn Astronomical Institute, 
Rijksuniversiteit Groningen, Landleven 12, 9717 AD, 
Groningen the Netherlands}

\author[0000-0002-6095-7627]{Rajeshwari Dutta}
\affiliation{IUCAA, 
Postbag 4, Pune 411007, Ganeshkind, India}

\author[0000-0002-4288-599X]{Celine Peroux}
\affiliation{European Southern Observatory, Karl-Schwarzschild-Str. 2, D-85748 Garching bei M\"unchen, Germany}
\affiliation{Aix Marseille Université, CNRS, LAM (Laboratoire d’Astrophysique de Marseille) UMR 7326, F-13388 Marseille, France}

\author[0000-0002-9946-4731]{Marc Rafelski}
\affiliation{Space Telescope Science Institute,
3700 San Martin Drive, MD 21218, Baltimore, USA}
\affiliation{Department of Physics and Astronomy, Johns Hopkins University, MD 21218, Baltimore, USA}

\author[0000-0002-4917-7873]{Mitchell Revalski}
\affiliation{Space Telescope Science Institute,
3700 San Martin Drive, MD 21218, Baltimore, USA}



\begin{abstract}
We detect \lya\ emission from a $\approx 5$~Mpc-long (comoving) portion of the cosmic web hosting an overdensity ($\delta \approx \overdens$) of \nlae\ \lya\ emitters (LAEs) at $z\approx 4$ within the MUSE Ultra Deep Field (MUDF), reaching an average surface brightness (SB) of $5\times 10^{-20}$~\sblcgs. This large-scale structure has an average SB similar to the filament across the two MUDF quasars at $z\approx 3.22$. However, deep multiwavelength data do not show a clear presence of active galactic nuclei, suggesting that the emission is mainly regulated by the underlying gas density. We find $\approx 0.2$~dex higher star formation compared to control samples and a remarkable predominance ($5/7$) of blue-peaked emission lines in the spectra of the embedded LAEs, indicative of favorable conditions for gas accretion. Lastly, we quantify the contribution of intragalactic gas to the \lya\ SB profile at large distances from LAEs. By studying samples of filaments detected in emission within diverse environments, we are finally gaining new insight into the physics of gas accretion within the cosmic web.  
\end{abstract}



\section{Introduction}\label{sec:intro}

The emergence of structures on scales $\gg 1~\rm Mpc$ is a distinctive prediction of a (cold) dark matter (DM) dominated Universe, in which matter aggregates in overdense sheets and filaments at the intersection of which DM halos form \citep{Klypin1983,Springel2005}. Although shaped by pressure, baryons follow the gravitational potential exerted by the DM \citep{Theuns1998,Bolton2017}, thus tracking the configuration of the large-scale structure (LSS)  and offering a test of theoretical predictions. 

Indirectly, the geometry of the LSS can be reconstructed by using galaxies as tracers of the underlying matter distribution. Over thirty years of wide-area galaxy surveys have provided exquisite statistical constraints on the nature of DM \citep{Davis1982,Tempel2014,Alam2017,Espinosa2022}. Studying the \lya\ forest in absorption in quasar spectra is an even more direct probe \citep{Rauch1998}. Through these techniques, the notion of cosmic web -- an intricate network of filaments connecting galaxies -- developed \citep{Bond1996}.

Obtaining direct images of the intergalactic gas within this web of filaments has been a long-standing goal, as mapping the morphology of the intergalactic gas with high resolution would augment the diagnostic power for studying the LSS and, hence, the nature of dark matter. However, the intrinsic low surface brightness (SB) of hydrogen ionized by the extragalactic background and the redshift-induced SB dimming \citep{Gould1996} made direct imaging of the cosmic web at $\approx 10^{-20}$~\sblcgs\ challenging for decades.

Focusing on $z\gtrsim 2$, the advent of sensitive integral-field spectrographs finally transformed this field. Initially, detections of filamentary tentacles extending from quasar host halos were reported \citep{Borisova2016,ArrigoniBattaia2018,Martin2019}, followed by the detection of emission patches aligned along filamentary structures extending for $>1~\rm Mpc$ in overdense regions, such as the SSA22 protocluster \citep{Umehata2019}. Ultradeep observations in the MUSE Extremely Deep Field \citep[MXDF;][]{Bacon2021} further unveiled evidence of widespread filamentary emission near star-forming galaxies, including a Mpc-scale filament connecting an active galactic nucleus (AGN) with \lya\ emitters (LAEs). 
A filamentary nebula has also been discovered in a compact LAE group \citep{banerjee_musequbes_2024}. 
Recently, a new ultradeep program in the MUSE Ultra Deep Field \citep[MUDF;][]{Fossati2019} provided a detailed view of a cosmic filament connecting two massive halos hosting quasars at $z\approx 3$, and measurements of its physical characteristics \citep{Tornotti2024}. In this work, we further extend the range of environments over which the cosmic web has been analyzed with detailed direct imaging and report a new detection of a filament connecting \nlae\ low-mass star-forming galaxies across $\approx 5$ comoving Mpc at $z\approx 4$ in the MUDF.


\begin{figure*}
\begin{tabular}{cc}
\begin{minipage}{0.62\textwidth} 
    \hspace*{-0.1\textwidth}
    \includegraphics[width=1.2\textwidth]{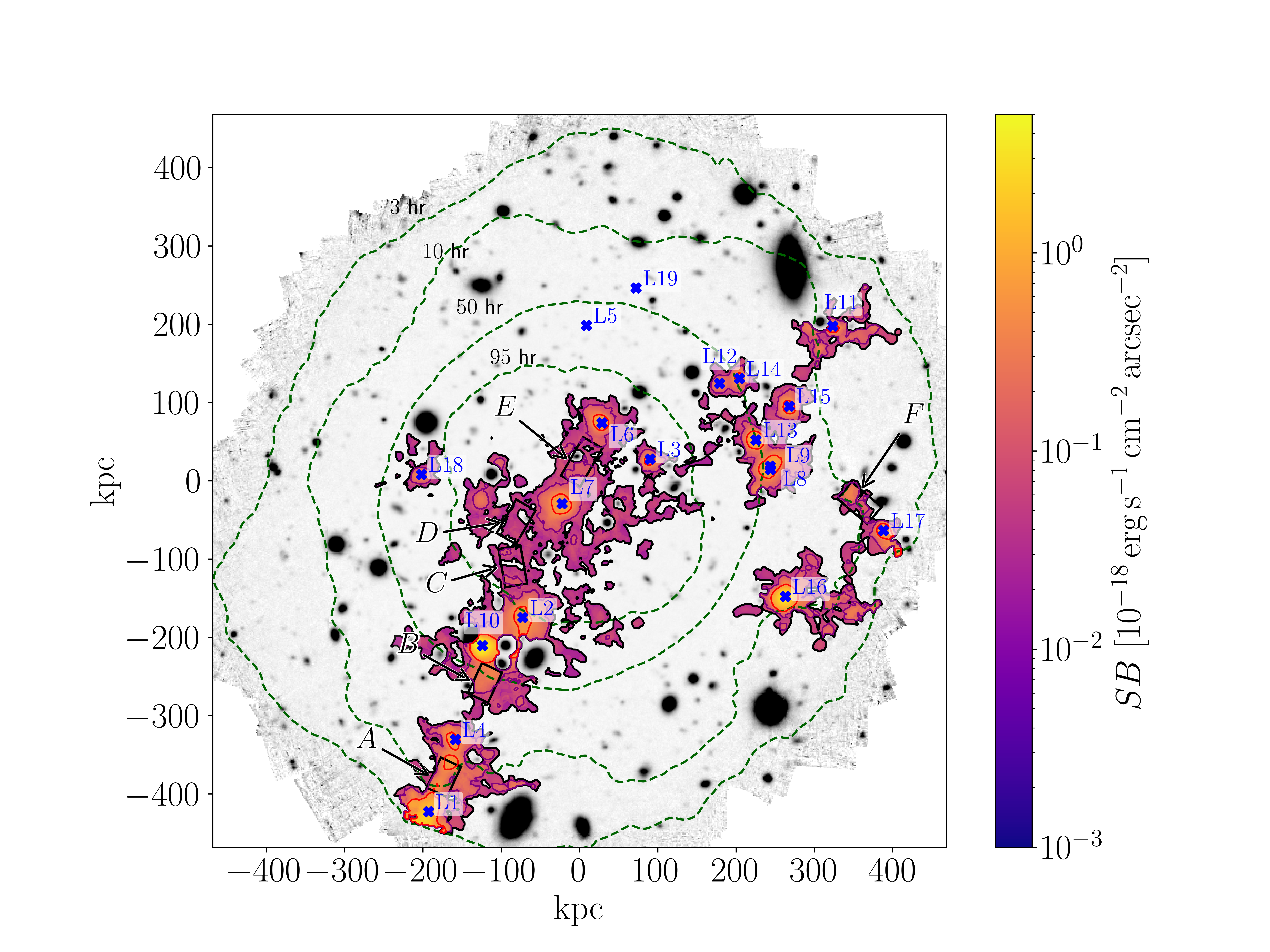} 
\end{minipage} &
\begin{minipage}{0.32\textwidth} 
    \vspace{0.01\textheight}
    \hspace*{0.01\textwidth}
    \includegraphics[width=0.95\textwidth]{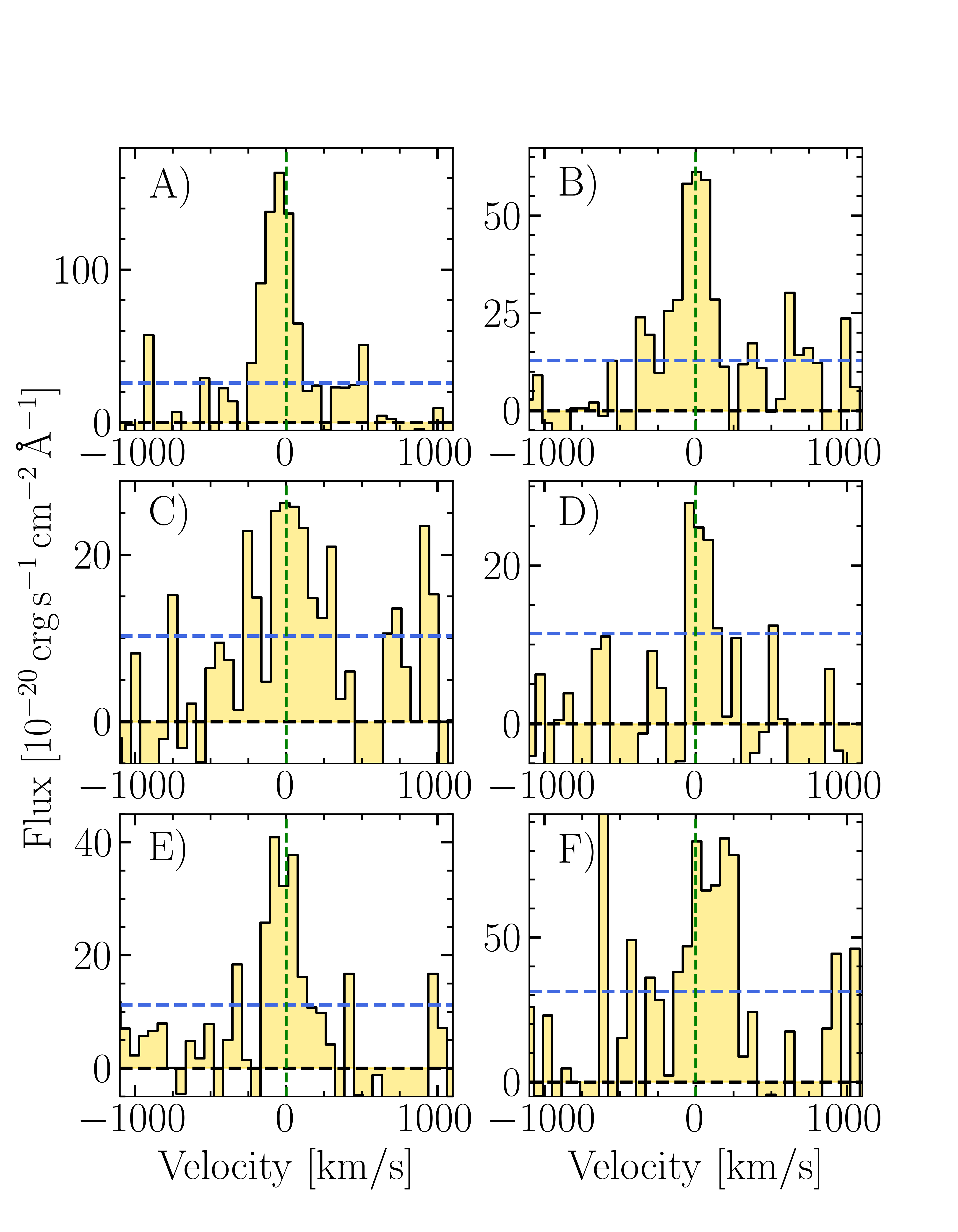} 
\end{minipage} \\
\end{tabular}
\label{fig:imgfilament}
\caption{{\it Left:} Extracted \lya\ image of a cosmic web filament connecting an overdensity ($\delta \approx \overdens$) of \nlae\ LAEs at redshift $z\approx 4.008$ obtained from the continuum-subtracted datacube. North is up and East to the left. The black contour is the detection limit at $S/N = 2$. The purple and red contours are at $1\times 10^{-19}$ and $5\times 10^{-19}$ \sblcgs.
The blue crosses indicate the LAEs identified at this redshift. L5 and L19 exhibit only compact emission unconnected to the extended signal and are not shown on this map. The background is a MUSE white-light image of the MUDF with continuum-detected sources visible in black and the dashed green contours defining the exposure time map of the field according to the labeled values. The black rectangles (A-F) are the extraction apertures of the spectra on the right panel. {\it Right:} Spectra of the \lya\ emission detected in the extraction windows, as marked on the left panel. These boxes focus on the spectral emission in regions of low $S/N$ and surface brightness, which are associated with the main filamentary structure.The horizontal dashed blue lines represent the $1\sigma$ uncertainty calculated from the region outside $\approx \pm 500~\rm km \, s^{-1}$ of the emission line centroid.}
\end{figure*}

\begin{figure}
\centering
\includegraphics[scale=0.35]{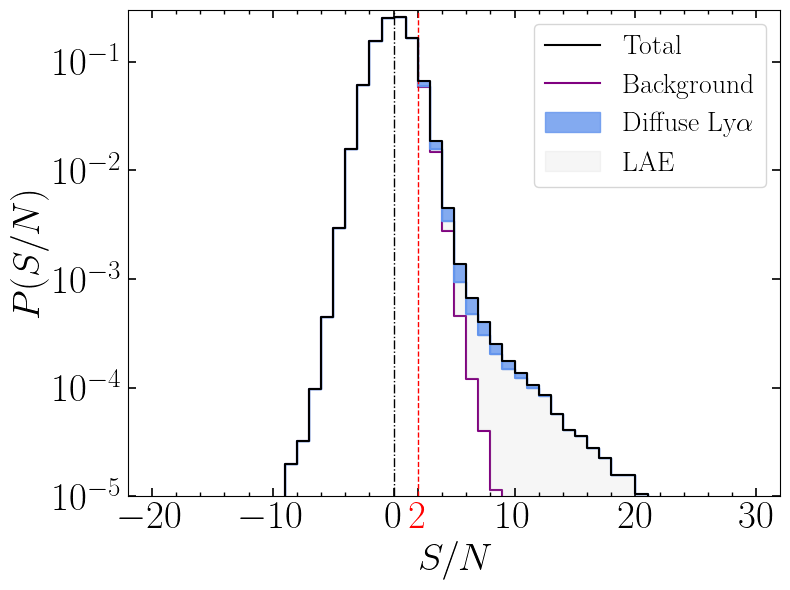}
\caption{The signal-to-noise distribution of the voxels in the datacube used for signal extraction at the redshift of the $z\approx 4.008$ LAE group. The purple line defines the distribution of the background. The gray region indicates the voxels associated with the emission from LAEs, while the blue region corresponds to the diffuse emission from the filaments (i.e, not overlapping in projection with SB levels of $4\times10^{-19}$ \sblcgs). The black dashed vertical line is the median of the background distribution ($\approx 0$) and the red dotted vertical line indicates the $S/N=2$ threshold.
}
\label{fig:SNRvox}
\end{figure}

\begin{figure*}
\centering
\includegraphics[scale=0.6]{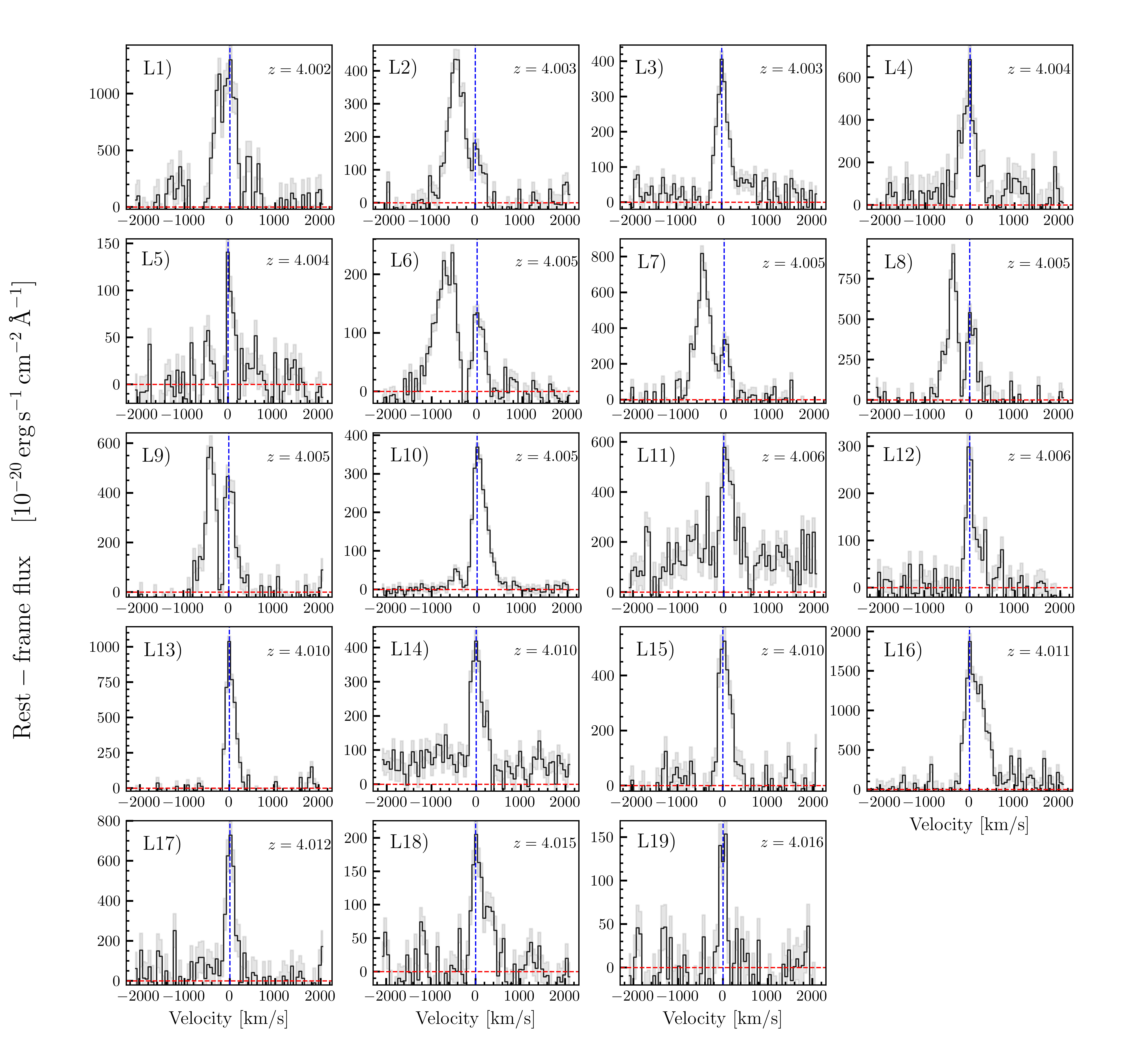}
\caption{Spectra of the \nlae\ LAEs associated with the analyzed group. The sources are sorted by redshift as labeled in Fig. \ref{fig:imgfilament}. The flux is rescaled to the rest-frame according to individual redshifts, determined as described in the text. The dashed blue vertical lines represent the zero velocity in each line, while the red dashed line is the zero flux reference. The gray shading represents the $1\sigma$ uncertainties in the flux values.}
\label{fig:laespec}
\end{figure*}

\section{Cosmic structures in the MUDF}\label{sec:mudf}

 The MUSE Ultra Deep Field is a 142-hour VLT/MUSE program (PID 110.A-0528) designed to map the connection between gas and galaxies at remarkable depth in a field hosting two bright quasars at $z\approx 3.22$. This field has also been the subject of extensive follow-up campaigns from the X-ray to the near-infrared using {\it Chandra}, {\it HST}, and ALMA. Details of observations and data reduction can be found in previous MUDF publications (\citealt{Lusso2019, Fossati2019, Revalski2023, Revaslki2024, Tornotti2024, Pensabene2024}). In this work, we rely on the full mean-combined datacube, presented in \citet{Tornotti2024}, that reaches a sensitivity of $3\times10^{-21}$~\fcgs~pix$^{-1}$ (at $\sim 5200$~\AA) in the central region. The cube has been prepared for analysis by subtracting the point spread function of the quasars and of the continuum sources through a non-parametric algorithm \cite[e.g.,][]{Borisova2016, ArrigoniBattaia2019}.

Two galaxy populations have been identified in the MUDF. The first population includes $3375$ sources, pinpointed by optical and NIR spectroscopy of continuum-detected galaxies in the deep F140W HST image. Details of this catalog are presented in \cite{Revalski2023}. The second population includes continuum-faint line emitters and, in particular, sources identified and classified by searching in the cube for \lya\ emission, as done in previous MUSE surveys \cite[for details, see][]{Fossati2021,galbiati_muse_2023}. For this work, we have produced a catalog, which we will present in detail in a forthcoming publication, containing $\approx 200$ LAEs at $z\gtrsim 3$ with $S/N > 7$ integrated over the line. Our catalog reaches \lya\ luminosities of $10^{41}~\rm erg~s^{-1}$ and is $50\%$ complete to $\approx 10^{41.5}~\rm erg~s^{-1}$ at $z\approx4$. The redshifts of the sources are assigned from the maximum of the \lya\ line. For double-peaked profiles without other non-resonant lines, the red peak is used to assign the redshift. Well-documented limitations due to resonant scattering can affect the redshift estimation, particularly in double-peak sources, with offsets of $\approx 200~\rm km~s^{-1}$  (e.g., \citealt{Verhamme2018, Muzahid2020, Galbiati2023}). 
The selection and properties of the groups are, nevertheless, not significantly affected.

\subsection{Identification of LAE groups}\label{sec:group}
The redshift distribution of the LAEs detected in the MUDF suggests the presence of regions where galaxies are grouped on scales of $\approx 1$~Mpc, a signpost that filaments may be present \citep{Bacon2021}. Therefore, we identify LAE groups using an algorithm based on the Friends-of-Friends approach
\cite[e.g.,][]{Huchra&Geller1982}. The galaxies are linked and considered group members if they are connected within linking lengths $\Delta R$ (a projected physical distance) and $\Delta v$ (a velocity in the redshift space). Using $\Delta R = 500$ kpc and $\Delta v = 400$~km~s$^{-1}$ (for the choice of parameters, see \citealt{Fossati2019}), we identify 32 systems of at least two or more galaxies. We checked that the identification of groups does not strongly depend on the choice of linking parameters and produces similar results even when these parameters are varied by $\pm 100$ kpc and $\pm 100~\mathrm{km~s^{-1}}$, respectively. We note, however, that the definition of linked systems is not unique, as other metrics can be adopted to connect galaxies in real and velocity space. Therefore, the groups we identify should be solely regarded as collections of galaxies to identify potential overdensities of LAEs where we can search for extended emission.
The complete LAE and group catalog will be presented in a forthcoming publication.

From this analysis, a group with a mean redshift $z\approx 4.008$ is notable for its richness, as it is composed of at least \nlae\ members (see Fig.~\ref{fig:imgfilament}). Its distribution in velocity space has a dispersion $\sigma \approx 245$~km~s$^{-1}$. All these galaxies are identified as LAEs, with only L4, L9, and L16 also showing continuum detection in HST/NIR imaging at the AB magnitude limit of $\approx 28$ in F140W \citep{Revalski2023}. No galaxies are detected in ALMA continuum observations \citep{Pensabene2024} to a flux limit at 1.2~mm of $\approx 0.1~$mJy~beam$^{-1}$. Compared to a mean galaxy population, this group corresponds to an overdensity of $\delta \approx \overdens$. Here, $\delta$ is defined as the number of observed LAEs divided by the expected number based on our entire catalog within a velocity window of $\approx 900~\rm km~s^{-1}$. 
To better explore the robustness of the overdensity estimate, we computed the expected number of LAEs assuming the field \lya\ luminosity functions of \cite{Thai2023} at $z\approx 2.9-4$ and $z\approx 4-5$. We integrate the luminosity function down to $\log(L_\mathrm{min}/\rm erg\,s^{-1})=41.5$, i.e. the lowest \lya\ luminosity in our group, correcting for our sample completeness. The effective comoving volume accounts for the area and depth of the survey and a line-of-sight $\Delta V = 2\sigma$ $\approx 5.8$ cMpc. We obtain estimates between $\delta \approx 18-29$, consistent with our measured $\delta \approx \overdens$. This range reflects the uncertainties in the estimates of $\delta$.
Moreover, this group is characterized by a spatial alignment of most of the LAEs in the south-east (SE) and north-west (NW) direction in a long and narrow region of thickness $\approx 150$~kpc, suggestive of a filamentary structure. The focus of this paper is to search and study the diffuse \lya\ emission from this region.

\begin{figure*}
\centering
\includegraphics[scale=0.45]{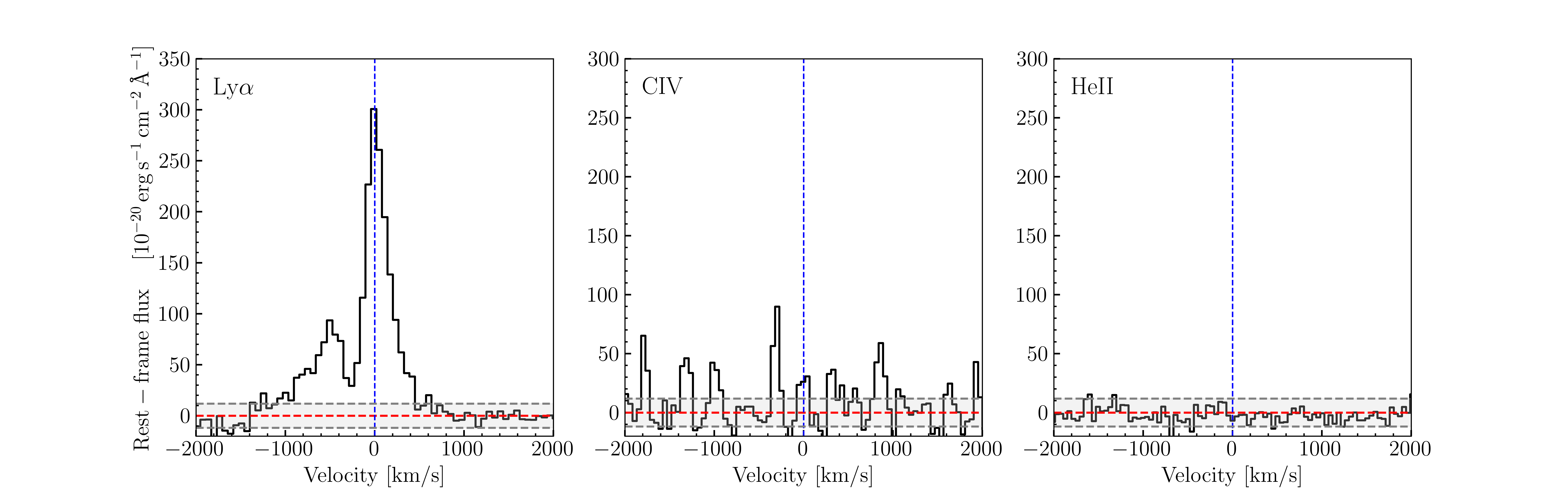}
\caption{Mean spectral stack for the \nlae\ LAEs associated with the analyzed group. From the left, the three panels show the spectral region associated with the \lya, \ion{C}{4} and \ion{He}{2} lines. The dashed blue vertical lines represent the expected zero velocity in each line, while the red dashed line is the zero flux reference. The grey-shaded region represents the $1\sigma$ noise estimated near each line ($\pm 3000$ km/s). Sharp peaks in the \ion{C}{4} region are associated with residual skylines in the cube.}
\label{fig:laespecstack}
\end{figure*}

\section{Imaging the extended \lya\ emission}\label{sec:filament}

At the mean redshift of this LSS, we search for extended \lya\ emission around the LAEs following a similar methodology to what is detailed in \cite{Tornotti2024}. Specifically, to extract extended emission we identify connected voxels ($>4000$) with a signal-to-noise ratio ($S/N$) above a threshold of 2, with a minimum number of $500$ spatial pixels.To increase the sensitivity at low SB, we apply a Gaussian kernel of $\sigma=4$ pixels ($0.8$~arcsec) in the spatial directions, masking all continuum sources obtained from the white-light image to avoid contamination from positive or negative residuals. We do not apply any smoothing kernel in the wavelength direction to preserve spectral resolution. Instead, we require at least $3$ adjacent spectral layers to be connected to avoid spurious detections. To identify connected pixels we use a Python code \citep[SHINE,][]{Fossati2025} which we made publicy available\footnote{\url{https://github.com/matteofox/SHINE}}. The code builds on an implementation of the algorithm by \citet{Rosenfeld1968} extended to 3D datasets using a decision tree algorithm known as Scan plus Array-based Union-Find \citep[SAUF,][]{Wu2005}, from the cc3d routine \citep{Silversmith2021}.
We then classify as emission associated with this structure all groups of voxels that have a median velocity within $1000$~km~s$^{-1}$ from the mean redshift of the LAE group. 
We have also verified that the morphology of the extended emission is robust with respect to the parameter choice. Extracting the signal with different spatial smoothing (3 pixels) and $S/N$ thresholds (up to 2.5) would still identify the same extended emission on large scales.
The detected signal is finally projected along the wavelength direction to obtain an extracted \lya\ image. The result is shown in Fig.~\ref{fig:imgfilament} (left) superimposed to the white-light image of the field. Emitters L5 and L19 are too compact and not connected to the extended signal to be represented on this map following this algorithm. Fig.~\ref{fig:SNRvox} shows instead the $S/N$ distribution of the extracted voxels, compared to the background where no signal is detected. 
From this analysis, we identify two distinct regimes of emission, separated by a surface brightness cut that we placed at $4\times 10^{-19}$~\sblcgs. The first regime corresponds to gas in the circumgalactic medium (CGM) of the LAEs, while the second is associated with diffuse gas that extends into low surface brightness region connecting the embedded LAEs within the filamentary structure. In Fig.~\ref{fig:SNRvox}, we illustrate this split in the $S/N$ distributions.
Only very deep data allow for the detection of such diffuse components beyond the patches of gas surrounding galaxies. 
To further confirm the real astrophysical nature of the detected signal, we extracted six spectra from regions surrounding the LAEs and their CGM (see rectangles A-F in Fig.~\ref{fig:imgfilament}). These regions were selected along the main filamentary structure, where the observations data are characterized by low surface brightness, and the use of a larger aperture would allow us to robustly test (i.e. at a higher S/N) the presence of signal that is more challenging to detect. \lya\ emission is also detected spectrally, ruling out the possibility that the extracted signal is an artifact of noise (Fig.~\ref{fig:imgfilament}, right).  

\section{Properties of a \lowercase{$z\approx 4$} cosmic filament}\label{sec:discussion}

The newly discovered MUDF filament is characterized by a main coherent structure that extends for $\approx 650~$kpc in the SE-NW direction ($\gtrsim 3$~Mpc comoving), with a thickness of $\approx 100~$kpc as visible on the map above $\approx 3\times 10^{-20}$~\sblcgs. 
Two secondary patches lie NW and W of the main structure. Overall, the ultra-deep MUDF observations image a portion of the cosmic web extending for at least $\approx 1~$Mpc ($\gtrsim 5$~Mpc comoving), making this one of the largest stretches of cosmic web imaged to date.
A bright patch of gas in the SE direction, at the edge of the field of view, points to a much more extended LSS that crosses the MUDF region. A range of SB is visible across the filament. The bright regions above $\approx 4\times 10^{-19}~$\sblcgs\ can be identified as the halo gas within $\approx 10-15~$kpc of LAEs (see below), as also demonstrated by the analysis of the radial profiles of LAEs with luminosities $\gtrsim 10^{41.5}~\rm erg~s^{-1}$ \cite[typical for our LAEs,][]{Wisotzki2018, Guo2024}.

At lower SB, reaching $\approx 3\times 10^{-20}~$\sblcgs, data reveal a more diffuse component that we associate with the intergalactic medium (IGM), with an average SB of $5\times 10^{-20}$~\sblcgs. Despite the very different environment of the filament reported by \cite{Tornotti2024} across two quasar hosts at $z\approx 3.22$, their average SB level is $8\times 10^{-20}$ \sblcgs, remarkably similar to the filament presented here. Considering the redshift-dependence $(1+z)^4$ of the SB dimming (a factor $2$ higher at $z\approx 4$), we conclude that this high-redshift filament is at least of comparable, if not higher, intrinsic brightness. Although interpreting the detailed origin of the emission mechanism is a difficult task and is the subject of extensive discussion in the literature \citep{elias_detecting_2020,byrohl_cosmic_2023,tsai_diffuse_2024}, the differential analysis of the two filaments provides new insight into the nature of the emission. First, we observe that the presence of two bright quasars does not appear to drastically affect the emergent SB, which is different from what is postulated in the optically thick regions \citep{Gould1996} inside the CGM. However, galaxies in the overdensity can still collectively contribute to the photon budget \citep{Umehata2019}. Moreover, a higher emissivity towards a higher redshift seems generally predicted in models due to the higher typical gas densities \citep{tsai_diffuse_2024} and the quadratic dependence on density of two fundamental emission mechanisms (collisions and recombinations).  Locally, high gas densities can also be encountered in galaxy overdensities due to galaxy interactions.
An extensive discussion, supported by modeling, of what powers this and other filaments is deferred to future publications. New observational constraints on the various emission mechanisms can be derived as more and more filaments are detected \citep{Bacon2021, Tornotti2024,banerjee_musequbes_2024}.

Considering the galaxy population found at the same redshift, we observe that the near-totality of LAEs is embedded within the central filament. The lack of continuum detection down to $\approx 28$~mag in the F140W image and to $\approx 0.1~$mJy~beam$^{-1}$ in the 1.2~mm map leads us to conclude that this filament hosts a population of low-mass, moderately star-forming, and dust-poor galaxies ($M_* \approx 10^8-10^9~$~M$_\odot$, star formation rate $SFR \approx 1-5$~M$_\odot$~yr$^{-1}$). The Ly$\alpha$ emission lines for these galaxies are shown in Fig.~\ref{fig:laespec}. Inspection of individual spectra rules out the presence of bright  \ion{C}{4} ($\lambda\lambda 1548, 1552$~\AA) and \ion{He}{2} lines ($\lambda 1640$~\AA), which, given their high ionization potential, are typically excited by AGNs. 

Moreover, stacking all the spectra (Fig.~\ref{fig:laespecstack}) results in a deep upper limit (with flux $\lesssim 6 \times 10^{-19}$~\flcgs\ for \ion{C}{4} and $\lesssim 3 \times 10^{-19}$~\flcgs\ for \ion{He}{2} lines in a $300$~km~s$^{-1}$ window). These upper limits translate in \ion{C}{4}/\lya\ of $\lesssim 2 \%$ and \ion{He}{2}/\lya\ of $\lesssim 1 \%$. Typical AGN spectra \citep{hainline_rest_2011} at $z\approx 3$ are characterized by ratios of $\approx 12-18 \%$ for \ion{C}{4}/\lya\ and $\approx 4-11 \%$ for \ion{He}{2}/\lya\ (depending on the equivalent width of \lya), higher than our limits. 
The nondetection of ALMA and X-ray sources \citep{lusso2023} further rules out the presence of heavily-obscured accreting black holes. We cannot exclude the presence of low-luminosity AGNs, which are likely to be present in this overdense region. Current deep and multiwavelength observations in the MUDF are, nevertheless, suggesting that the ionizing flux is not dominated by an abundant population of particularly bright and active black holes, as also seen in the filament reported at $z\approx 3.6$ by \citet{banerjee_musequbes_2024}.
Moreover, we cannot exclude the potential presence of undetected galaxies, which may contribute to the extended \lya\ emission, as suggested, e.g., by \cite{Bacon2021}.

\begin{figure}
\centering
\includegraphics[scale=0.45]{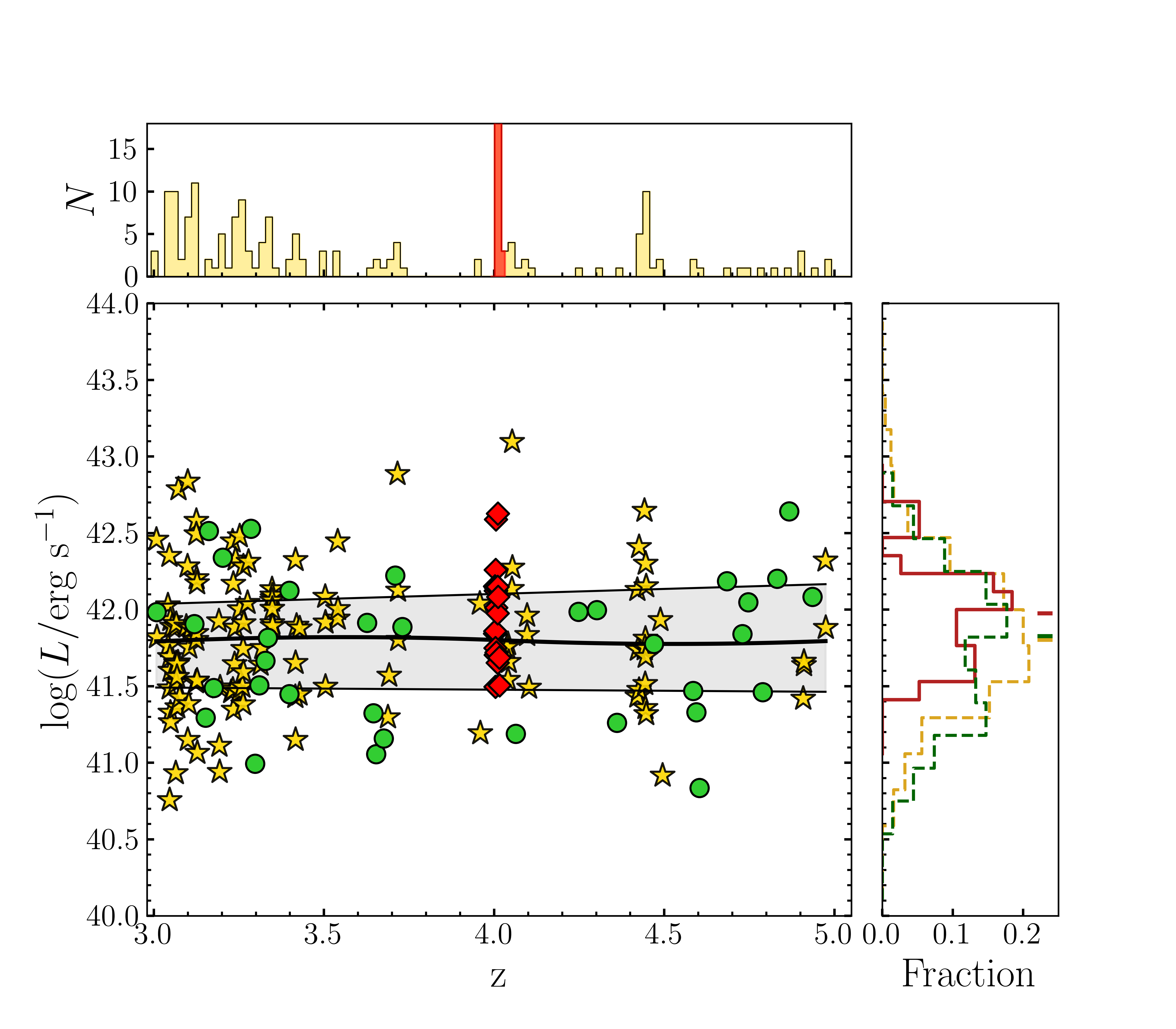}
\caption{Comparison of the \lya\ luminosity of galaxies in the $z\approx 4.008$ filament (red diamonds) and in a control sample, which we divide in galaxies in groups (125, yellow stars) or more isolated systems (34, green circles). The solid black line and the associated gray region represent the smoothed median, $25 \rm th$ and $75 \rm th$ log-luminosity percentile regression curves for the control sample as a function of redshift, obtained with a B-splines (COBS) smoothing algorithm (\citealt{Ng&Maechler2007, Ng&Maechler2024}). The luminosity distributions, marginalized in redshift, are shown as histograms in the right panel. The top panel shows the redshift distribution of the sample, with the red solid line marking the filament redshift.}
\label{fig:laecontrol}
\end{figure}

Among the LAEs in this group, seven sources (L2 and L5-10; $\approx 37$~percent of the total) exhibit a double peak profile according to visual classification. This fraction is consistent with the incidence of multipeaked \lya\ lines at similar redshifts and luminosities \citep{Kulas2012, Vitte2024}. Notably, all the double-peaked profiles but two (L5, L10) show a marked blue peak compared to the red peak, a distinctive feature attributed to infalling gas onto these galaxies \citep{dijkstra_ly_2006,verhamme_3d_2006}. Typically, the fraction of galaxies in which the blue peak dominates over the red peak is below $20-40$~percent in comparable datasets \citep{Kulas2012, Vitte2024}. With an incidence of 5/7 of high-confidence brighter blue peaks, we conclude that galaxies embedded in this filament may be experiencing favorable conditions for gas accretion. We also explore the kinematics of the gas in the filament more closely. Despite the depth of the data, a pixel-by-pixel analysis is prohibitive because of the faint SB. We, therefore, proceed by examining the shifts in the velocity centroid of the \lya\ line in the regions highlighted in Fig.~\ref{fig:imgfilament}. No clear velocity gradient appears beyond the tolerance imposed by using a resonant line, $\approx 200~$km~s$^{-1}$. Futhermore, no noticeable velocity gradient emerges when considering the LAE redshifts. Thus, the filament appears to have a relatively uniform kinematic structure, as also seen in \citet{Tornotti2024}. We speculate that this commonality might be related to a preference in detecting emssion from face-on filaments for which we cannot robustly measure kinematics. The expectation supported by numerical simulations is that gas flows along the filaments with typical velocities of a few hundred kilometers per second, to reach the galaxies and sustain the formation of new stars \citep{dekel_cold_2009,martin_multi-filament_2019}. As the sample of detected filaments grows, combinations of observations and simulations will shed further light on accretion through filaments.

The fact that a significant fraction of galaxies embedded in a gas-rich filament also display spectral signatures of infalling gas opens the question of whether they are experiencing more active star formation. We investigate this in Fig.~\ref{fig:laecontrol}, where we show the \lya\ luminosity of the members of this structure with the LAE population detected in the MUDF between $z\approx 3-5$. Across this redshift, the intrinsic properties of LAEs are weakly evolving, and cosmic time dependence is negligible in this comparison. The \nlae\ LAEs at $z\approx 4.008$ are generally distributed as the control sample of 159 galaxies (defined as the LAEs not belonging to this group), with a $\approx 0.2$~dex offset at higher \lya\ luminosity distribution (the median \lya\ in this sample is $\log(L/\mathrm{erg~s^{-1}}) \approx 42$). To investigate the significance of this difference, we randomly draw from the control sample $500$ sub-samples of \nlae\ galaxies, and we obtain a median value of $\log(L/\mathrm{erg~s^{-1}}) = 41.80^{+0.10}_{-0.15}$. Thus, a marginal \lya\ offset is present at $2\sigma$ significance. 
Splitting the control sample in group LAEs, which could trace a similar environment to the filament, and more isolated systems yields comparable results (yellow and green histograms in Fig.~\ref{fig:laecontrol}).
This analysis hints at a significant star formation activity within the filament, as expected for regions of enhanced gas accretion \citep{dave_analytic_2012,lilly_gas_2013}. However, additional information (e.g., on the stellar mass or UV-based star formation rates) is required to investigate this trend more robustly.  

\begin{figure}
\centering
\includegraphics[scale=0.45]{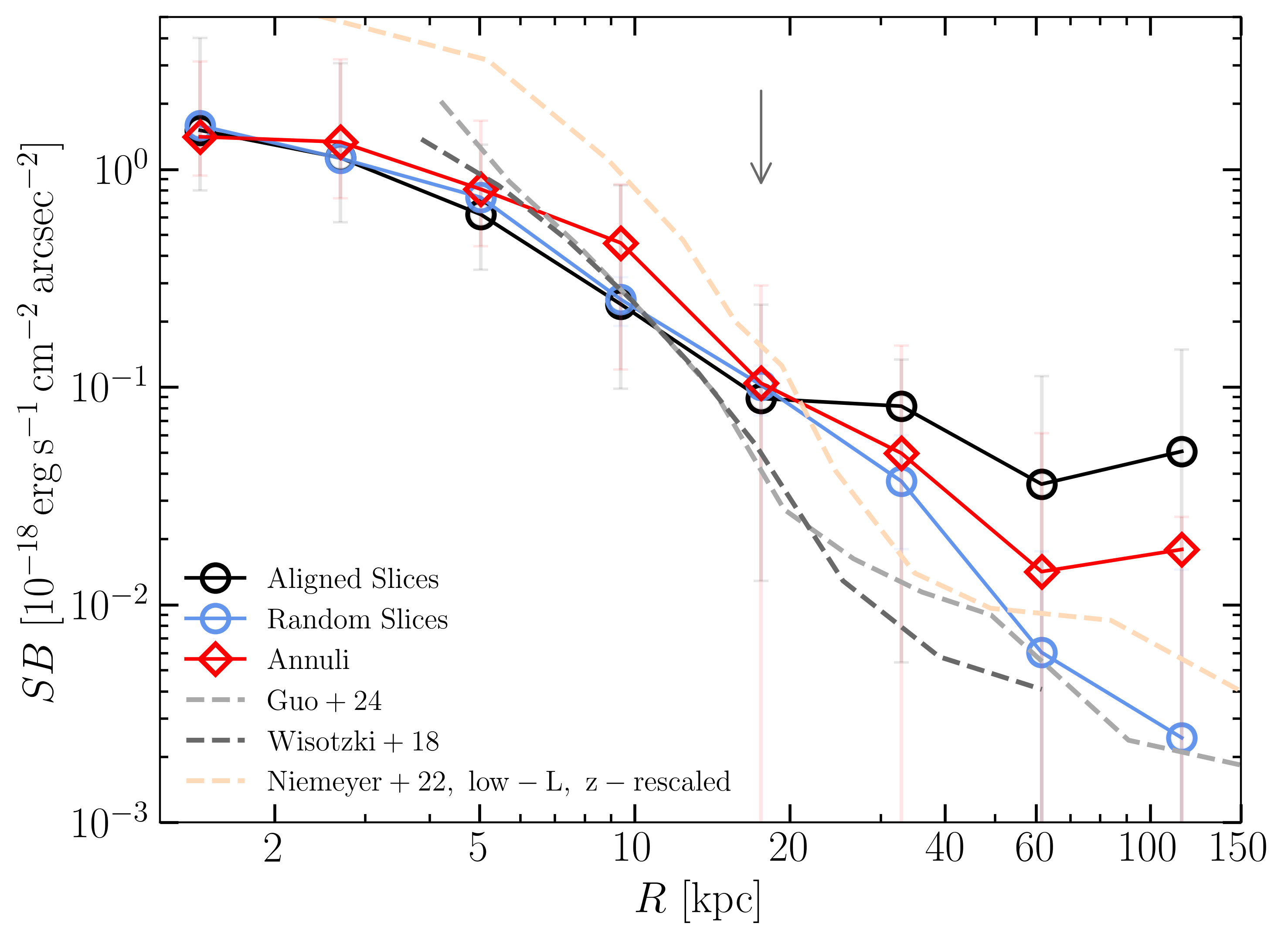}
\caption{Median SB profile of LAEs extracted along slits aligned with the filament emission (black), randomly-oriented slits (blue), and annuli (red). The vertical arrow marks the inflection point we identified at $\approx 20~$kpc. Profiles from large samples of LAEs from the literature are also shown with black, grey and orange dashed lines.}
\label{fig:laesbprofiles}
\end{figure}

The detection of a considerable sample of LAEs within \lya\ emitting gas provides a helpful laboratory to study how the CGM of galaxies connects to the IGM. To include all the flux, we construct pseudo narrow-band images of 30~\AA\ bandwidth for each LAE, centering on the peak wavelength of the \lya\ line. 
We extract SB profiles at the top of each LAE along a slit in the direction that aligns with the detected filament.
The slit width varies in the interval $5-40$~kpc moving outward,
while the box length used to sample the profile along the slit is logarithmically spaced. This choice of varying width mimics the adaptive nature of the annuli, which sample the inner regions at higher resolution. The result is shown in Fig.~\ref{fig:laesbprofiles}. We observe a steepening of the median profile between $\approx 10-20$~kpc, followed by a plateau of almost constant SB at $4-6\times 10^{-20}$~\sblcgs\ at $\gtrsim 20~$kpc. We interpret the inflection point at around $\approx 20$~kpc as a signpost of the transition between the emission coming from the gas associated with the halo (i.e., the CGM) and the surrounding gas in the IGM. A similar separation is also seen in the filament by \citet{Tornotti2024}, who reported evidence of a break in the SB profile around the virial radius of halos embedded inside the cosmic web. In fact, this size is comparable to the virial radius of $10^{11}~\rm M_\odot$ halos at $z\approx 4$, which is $R_{\rm vir}\approx 30$~kpc ($R_{\rm vir}\approx 20$~kpc for masses of $10^{10.5}~\rm M_\odot$; \citealt{HerreroAlonso2023,Guo2024}).
Beyond this empirical separation, the resulting SB profile can be shaped by the underlying density distribution and/or the emission mechanisms. Separating these contributions is, however, a task we defer to future work. Moreover, within the halo and in such dense environments, outflows and tidal interactions can affect the density distribution and shape the radiation field, influencing the \lya\ emission (e.g., \citealt{Umehata2019, Cantalupo2014}).

We can also use the direct detection of this filament to compare with the results of stacking experiments conducted on large LAE samples \citep{leclercq_muse_2017,Wisotzki2018,LujanNiemeyer2022,Guo2024}. In these stacks, a halo profile is identified up to $\approx 40-50~$kpc, at which point a flattening to large distances ($\gg 100~$kpc) is observed. This change in slope can be statistically attributed to the transition between the CGM and the IGM. However, in stacks, this interpretation is complicated by geometric dilution, i.e., the stacking of signals aligned along filaments or across empty regions, further complicated by the range of LAE virial radii. Moreover, models that capture the effect of (unresolved) nearby halos \citep{byrohl_cosmic_2023,Bacon2021} seem to reproduce the SB plateau well with a two-halo term. We directly test the first scenario by stacking SB profiles of LAEs extracted from our map at random orientations, both in slits (blue line in Fig.~\ref{fig:laesbprofiles}) and in annuli (red line) to mimic the unknown geometry of the undetected filaments. To increase the significance of the measurement in the case of random slits, we combine the sample of \nlae\ LAEs 500 times, each time selecting a random orientation. 

A comparison of these different profiles (Fig.~\ref{fig:laesbprofiles}) reveals, in the inner region, a substantial agreement between the aligned and random slits.
This is explained by the fact that the regions at $\lesssim 20~$kpc are filled by isotropic emission. At larger radii, the SB profile obtained with random slits steeply declines, as the apertures often select regions devoid of emission. 
More interesting is the case of the annuli, where a plateau ascribed to the filament emission is still visible, although with a lower average SB because of the geometric signal dilution. This exercise highlights that stacking analysis of large samples can reveal an emission signal from the cosmic web. However, the recovered value of SB does not necessarily reflect the mean SB of the filaments because of the presence of an unknown fraction of empty regions.
Examples of stacked profiles from MUSE data in large samples at $3 \lesssim z \lesssim 4$ from \citet{Wisotzki2018} and \citet{Guo2024} are shown in Fig.~\ref{fig:laesbprofiles} (dashed lines). Additionally, the rescaled and dimming-corrected profile of the low luminosity ($\log(L_\mathrm{Ly\alpha}/\rm erg\,s^{-1}) < 43$) sub-sample at $z \approx 2.5$, derived from the Hobby-Eberly Telescope Dark Energy Experiment (HETDEX) data by \cite{LujanNiemeyer2022}, is also reported. Compared with our median profile in annuli, we observe substantial agreement in the inner regions, except for the \cite{LujanNiemeyer2022} sub-sample composed of sources with higher luminosity ($\log(L_\mathrm{Ly\alpha}/\rm erg\,s^{-1})>42.4$)
than the median of our sample $\log(L_\mathrm{Ly\alpha}/\rm erg\,s^{-1})\approx 42$. However, the literature profiles decrease much below our result for $\approx 20$~kpc, and a flattening is observed only at $\approx 80-100~$kpc. A possible interpretation of this difference is that the filament we are detecting is generally brighter than the average population, a reasonable conclusion for the first detections of any class of objects that are often the tip of the iceberg. In more typical cases, the outer CGM (between $\approx 20-60$~kpc) can still outshine the filament emission that is revealed only at greater distances.

\section{Summary and conclusions}\label{sec:conclusion}

Using ultra-deep MUSE observations within the MUDF,
we report the discovery of a $\approx 5~$Mpc (comoving) portion of a $z\approx 4.008$ cosmic-web filament detected in \lya\ emission within a $\delta \approx \overdens$ overdensity of \nlae\ LAEs. We jointly analyze the emission properties of the diffuse gas in the filament and of the embedded LAEs, and we report the following findings.

\begin{itemize}
    \item[--] The diffuse emission outside of the CGM of individual galaxies reaches SB levels of $\approx 3\times 10^{-20}~$\sblcgs\ and outlines an LSS that extends for $\approx 1~$Mpc (proper), with thickness of $\approx 100$~kpc.  
    \item[--] Despite the substantially different galaxy environment, the basic emission properties of the filament resemble the one detected across the two luminous MUDF quasars at $z\approx 3.22$. The lack of evident AGNs in this overdensity suggests that the underlying gas density, rather than the radiation field, regulates the emission properties on IGM scales, but galaxies in the overdensity can still contribute collectively to a significant photon budget. 
    Interactions among galaxies may also contribute to higher local densities. 
    \item[--] Most (5/7) of the double-peaked \lya\ profiles of the embedded LAEs display a dominant blue peak, a signature that can be explained by an increased incidence of gas accretion. LAEs in this filament also show a $2\sigma$ excess of \lya\ emission compared to control samples, hinting at higher star-formation activity.
    \item[--] By stacking the LAE SB profile in slits aligned with the filament, we identify the
    transition between the CGM and IGM at a radius of $\approx 20~$kpc, which is comparable to the virial radius expected for LAE hosts. Moreover, we show that the flattening at large radii in stacked profiles from literature samples can be interpreted as emission arising from filaments fainter than the one we detect, albeit modulated by the filling factor of the emitting regions.  
\end{itemize}

Current detections with surface brightness approaching $5\times 10^{-20}~$\sblcgs, including our new example, arise from searches conducted around galaxy groups \citep{Umehata2019,Bacon2021, Tornotti2024,banerjee_musequbes_2024}. A link between galaxy overdensities and cosmic-web emission is thus emerging.
The range of environments and the richness of the groups probed vary substantially, from rare protoclusters like SSA22 to more common LAE groups. We conclude that filaments emitting at $\approx 5\times 10^{-20}~$\sblcgs\ could be somewhat ordinary. Medium-depth programs sensitive to extended emission with $\gtrsim 10^{-20}~$\sblcgs\ have the potential of yielding the first significant samples of cosmic web images.  

An open question that emerges is what sets the observed surface brightness level across various environments. 
Our study shows a possible way forward:
the comparative analysis of \lya\ imaging in filaments within various environments jointly with the properties of embedded galaxies. As larger samples of directly-imaged filaments become available, the homogeneous study of the diffuse emitting gas and the cospatial galaxy population will finally constrain the frequency and physical properties of the cosmic web with solid empirical evidence. A new era of detailed studies of cosmic filaments in emission has finally opened.   

\begin{acknowledgments}
This project has received funding from Fondazione Cariplo (grant No 2018-2329) and is supported by the Italian Ministry for Universities and Research (MUR) program ``Dipartimenti di Eccellenza 2023-2027'' within the Centro Bicocca di Cosmologia Quantitativa (BiCoQ). This work is based on observations collected under ESO programme ID 1100.A-0528. PD acknowledges support from the NWO grant 016.VIDI.189.162 (``ODIN") and warmly thanks the European Commission's and University of Groningen's CO-FUND Rosalind Franklin program. This research made use of Astropy, a community-developed core Python package for Astronomy (\citealt{Astropy2013, Astropy2018, Astropy2022}, NumPy (\citealt{Harris2020}), SciPy (\citealt{Virtanen2020}), Matplotlib (\citealt{Hunter2007}).
\end{acknowledgments}

%

\vspace{20mm}
\facility{VLT(MUSE)}







\bibliography{ms}{}
\bibliographystyle{aasjournal}



\end{document}